\PassOptionsToPackage{unicode}{hyperref}
\PassOptionsToPackage{hyphens}{url}
\documentclass[12pt]{article}
\usepackage{amsmath}
\usepackage{graphicx,psfrag,epsf}
\usepackage{enumerate}
\usepackage[]{natbib}
\usepackage{textcomp}

\newcommand{\blind}{0}

\providecommand{\tightlist}{%
  \setlength{\itemsep}{0pt}\setlength{\parskip}{0pt}}

\IfFileExists{bookmark.sty}{\usepackage{bookmark}}{\usepackage{hyperref}}
\IfFileExists{xurl.sty}{\usepackage{xurl}}{} 
\hypersetup{
  pdfkeywords={American football, defensive linemen, multilevel
model, player tracking data},
  hidelinks,
  pdfcreator={LaTeX via pandoc}}

\begin{document}

\date{July 31, 2023}

\def\spacingset#1{\renewcommand{\baselinestretch}%
{#1}\small\normalsize} \spacingset{1}


\if0\blind
{
  \title{\bf Here Comes the STRAIN: Analyzing Defensive Pass Rush in
American Football with Player Tracking Data}

  \author{
        Quang Nguyen \\
    Department of Statistics \& Data Science\\
Carnegie Mellon University\\
     and \\     Ronald Yurko \\
    Department of Statistics \& Data Science\\
Carnegie Mellon University\\
     and \\     Gregory J. Matthews \\
    Department of Mathematics and Statistics\\
Center for Data Science and Consulting\\
Loyola University Chicago\\
      }
  \maketitle
} \fi

\if1\blind
{
  \bigskip
  \bigskip
  \bigskip
  \begin{center}
    {\LARGE\bf Here Comes the STRAIN: Analyzing Defensive Pass Rush in
American Football with Player Tracking Data}
  \end{center}
  \medskip
} \fi

\begin{abstract}
In American football, a pass rush is an attempt by the defensive team to
disrupt the offense and prevent the quarterback (QB) from completing a
pass. Existing metrics for assessing pass rush performance are either
discrete-time quantities or based on subjective judgment. Using player
tracking data, we propose STRAIN, a novel metric for evaluating pass
rushers in the National Football League (NFL) at the continuous-time
within-play level. Inspired by the concept of strain rate in materials
science, STRAIN is a simple and interpretable means for measuring
defensive pressure in football. It is a directly-observed statistic as a
function of two features: the distance between the pass rusher and QB,
and the rate at which this distance is being reduced. Our metric
possesses great predictability of pressure and stability over time. We
also fit a multilevel model for STRAIN to understand the defensive
pressure contribution of every pass rusher at the play-level. We apply
our approach to NFL data and present results for the first eight weeks
of the 2021 regular season. In particular, we provide comparisons of
STRAIN for different defensive positions and play outcomes, and rankings
of the NFL's best pass rushers according to our metric.
\end{abstract}

\noindent%

\vfill

\newpage
\spacingset{1.9} 

\newpage
\spacingset{1.45}

\hypertarget{sec:introduction}{%
\section{Introduction}\label{sec:introduction}}

In recent years, tracking data have replaced traditional box-score
statistics and play-by-play data as the state of the art in sports
analytics. Numerous sports are collecting and releasing data on player
and ball locations on the playing surface over the course of a game.
This multiresolution spatiotemporal source of data has provided
exceptional opportunities for researchers to perform advanced studies at
a more granular level to deepen our understanding of different sports.
For complete surveys on how tracking data have transformed sports
analytics, see \citet{Macdonald2020Recreating}, \citet{Baumer2023Big},
and \citet{Kovalchik2023Player}.

In an attempt to foster analytics and innovate the game, the National
Football League (NFL) introduced their player tracking system known as
Next Gen Stats in 2016 \citep{nfl2023ngs}. Next Gen Stats uses radio
frequency identification (RFID) chips placed in players' shoulder pads
(and in the ball) to collect data at a rate of 10 frames per second. The
data capture real-time on-field information such as locations, speeds,
and accelerations of all 22 players (and the football). While these data
were initially only available for teams, media, and vendors, in December
2018 the NFL launched the inaugural edition of their annual Big Data
Bowl competition \citep{nfl2023big}.

The first Big Data Bowl led to several contributions largely focused on
offensive performance evaluation. For example, one group of finalists
introduced an approach for modeling the hypothetical completion
probability of a pass aiding in the evaluation of quarterback (QB)
decision making \citep{deshpande2020expected}. The winners of the
inaugural Big Data Bowl focused on identifying receiver routes via
clustering techniques \citep{chu2020route} and convolutional neural
networks \citep{sterken2019routenet}. Along with the competition
entries, the public release of NGS data allowed researchers to tackle a
variety of other problems such as revisiting fourth down decision making
\citep{lopez2020bigger}, annotating pass coverage with Gaussian mixture
models \citep{dutta2020unsupervised}, and introducing a continuous-time
framework to estimate within-play value \citep{yurko2020going}. Since
its inception, the Big Data Bowl has chosen a different theme each year
leading to new insight about evaluating different positions such as
running backs, defensive backs, and special teams. The 2023 edition of
the NFL Big Data Bowl asked participants to evaluate linemen on passing
plays \citep{Howard2023NFL}.

\begin{figure}[t]

{\centering \includegraphics{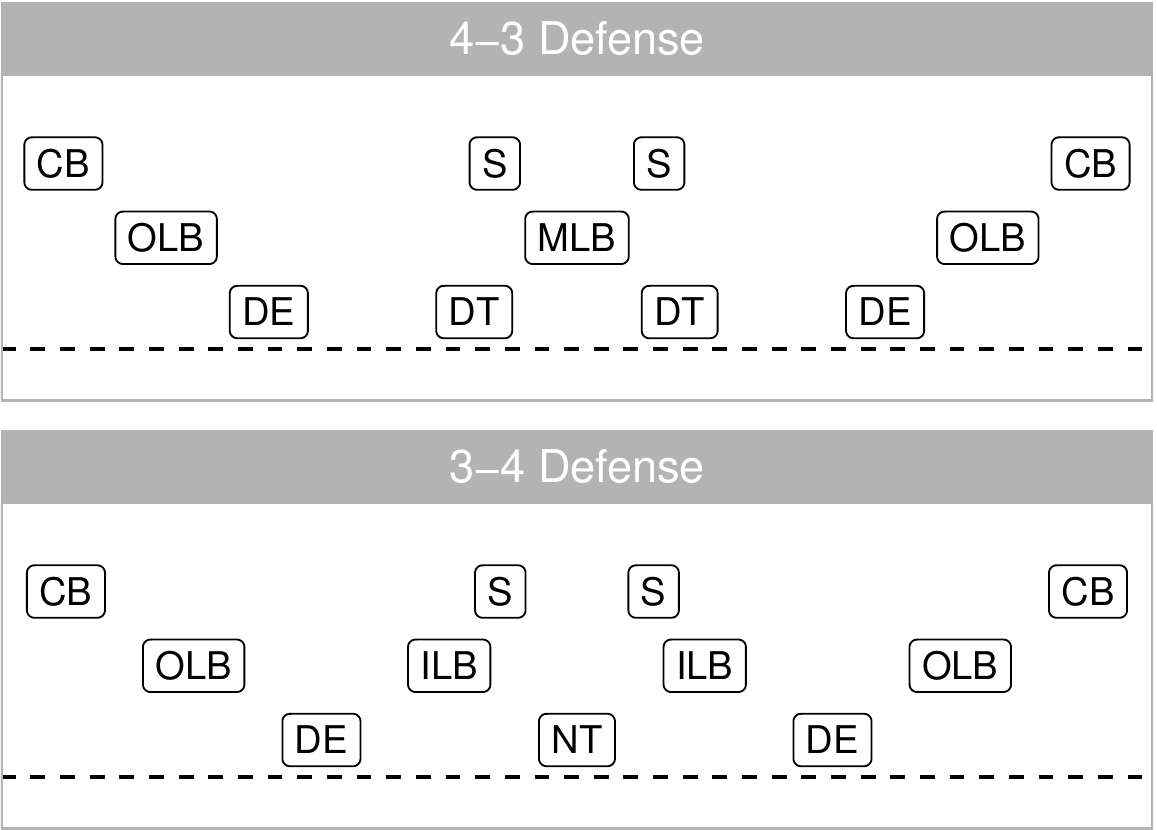} 

}

\caption{Two common defensive alignments in football: 4-3 defense (top) and 3-4 defense (bottom). The dashed line represents the line of scrimmage separating the defense and offense. Defensive tackles (DT), defensive ends (DE), and nose tackles (NT) primarily rush the QB on passing plays and attempt to stop the ball carrier as quickly as possible on running plays. Outside linebackers (OLB) and inside linebackers (ILB) usually play directly behind the defensive line and are involved in defending against passing and rushing plays. Cornerbacks (CB) and safeties (S) generally involve in defending against passing plays.}\label{fig:formation}
\end{figure}

Our focus of this manuscript is specifically on measuring the
performance of defensive linemen in the NFL. There are two main types of
defensive linemen in American football: defensive tackles and defensive
ends. Typically, these positions are located within the interior of the
line and along the edges, respectively; see Figure \ref{fig:formation}
(top) for example formation with defensive tackles and defensive ends.
The primary purpose of both positions is to rush the QB on passing
plays, with defensive ends displaying superiority in observed pass
rushing ability \citep{Eager2018nfl}. Additionally, within defensive
tackles there are nose tackles who directly line up across from the ball
at the line of scrimmage; see Figure \ref{fig:formation} (bottom) for
example defensive scheme with a nose tackle. NFL teams often employ
either one nose tackle or two defensive tackles on the interior with
defensive ends along either side of the defensive line. Besides
defensive lineman, other positions may attempt to rush the QB on
blitzing plays such as outside linebackers, interior linebackers, and
potentially members of the secondary (cornerbacks, free safeties, and
strong safeties) whose primary role is pass coverage. Note that apart
from the formations shown in Figure \ref{fig:formation}, defensive
linemen can have the flexibility to line up differently. For example, a
defensive end, depending on the opposing matchup, may not necessarily be
positioned toward the outside of the line of scrimmage.

In this work, using data made available in the Big Data Bowl 2023, we
present a novel approach to measure the performance of pass rushers.
Relative to other aspects of American football, such as quarterback
evaluation \citep{burke2019deepqb, reyers2021quarterback}, the
literature on evaluating pass rushers is scarce. Below, we provide a
brief overview of existing pass rush metrics.

\hypertarget{previous-pass-rush-metrics}{%
\subsection{Previous Pass Rush
Metrics}\label{previous-pass-rush-metrics}}

Table \ref{tab:metrics} gives a summary of existing football metrics for
pass rush. We now highlight what these quantities describe as well as
their limitations.

Perhaps the most commonly-known statistics for evaluating defensive
linemen on pass rush plays are sacks, hits, and hurries, which are
discretely observed at the play-level. Officially tracked by the NFL
since 1982, a sack is recorded when a defender tackles the QB behind the
line of scrimmage before the QB releases a pass. Other traditional box
score statistics such as hits and hurries are collected by various
outlets. A hit is a collision between a defender and the opposing team's
quarterback after the quarterback makes a throw. A hurry represents an
instance when a defender successfully disrupts without necessarily
making direct contact with the QB and forces the QB to throw the
football earlier than expected. These are all simple binary measures of
pass rush outcome for any given play. However, for plays that do not
result in the aforementioned outcomes (in particular, sack), there are
still many intermediate defensive actions on the field within the play
that are valuable and can be considered positive achievements.

In addition, the sum of sacks, hits, and hurries is often defined as
pressures. This is better than the individual counts to some extent, but
suffers from problems of subjectivity (e.g., whether there is an actual
hurry or not). Pro Football Focus (PFF) defines a metric called
pass-rush productivity, which is a minor modification from the
aforementioned pressures metric (see Table \ref{tab:metrics}). In
particular, pass-rush productivity gives twice as much weight to a sack relative to hurries
and hits, which is a small upgrade to pressures. However, the choice of
weights is ad-hoc and still only considers binary outcomes, similar to
the shortcomings of previous metrics.

\begin{table}
\caption{A summary of previously-existing pass rush metrics. \label{tab:metrics}}
\centering
\begin{tabular}{p{0.27\textwidth}p{0.67\textwidth}}
\hline
Metric & Description \\ 
\hline
Sacks & A defender tackles the QB behind the line of scrimmage before a QB throw \\
Hits & A defender tackles the QB behind the line of scrimmage after a QB throw \\
Hurries & A defender pressures the QB behind the line of scrimmage forcing the QB to throw the ball sooner than intended \\
Pressures & $\text{Hurries} + \text{ Hits } + \text{ Sacks}$ \\
Pass-Rush Productivity & $\displaystyle \frac{(\text{Hurries} + \text{Hits})/2 + \text{Sacks}}{\text{ Pass Rush Snaps}}$\\
Time In Pocket & Time (in seconds) between ball snap and throw or pocket collapse for a QB \\
NGS Get Off & Average time (in seconds) required for a defender to cross the line of scrimmage after the ball snap\\
Pass Rush Win Rate & Rate at which pass rusher beats pass block within 2.5 seconds after ball snap\\
\hline
\end{tabular}
\end{table}

More recently proposed metrics such as time in pocket, NGS get off
\citep{Hermsmeyer2021nfl}, and pass rush win rate
\citep{Burke2018Created} are substantial improvements over the less
sophisticated counting statistics, but nevertheless are still imperfect.
Time in pocket refers to how long a QB can operate within the protected
space behind the offensive line, known as the pocket. However, this
measure is highly context-dependent, as it can be influenced by a number
of factors such as the defensive scheme or type of passing route. NGS
get off is an aggregated statistic, illustrating how quickly a defender
can get past the line of scrimmage after the snap on average. Pass rush
win rate is created using player tracking data, which is at a more
granular level than previous measures. It demonstrates whether a pass
rusher is able to beat their blocking matchup before a fixed time from
the snap (2.5 seconds as chosen by ESPN). However, this depends on the
rather arbitrary time threshold used to define a pass rush win. Besides,
once a cutoff is chosen, pass rush win rate converts continuous data to
a win-loss indicator, becoming dichotomous like most of the metrics
discussed above.

\hypertarget{previous-research-on-football-linemen}{%
\subsection{Previous Research on Football
Linemen}\label{previous-research-on-football-linemen}}

The peer-reviewed literature on measuring the performance of football
linemen (either offensive or defensive) is scant.
\citet{AlamarGould2008} find an association between pass completion rate
and successful pass blocking by offensive linemen. The data for this
study are collected for the first three weeks of the 2007 NFL season,
manually recording whether a lineman holds a block and the time it took
for the quarterback to throw the football. \citet{AlamarGoldner2011} later
follow up by using manually-tracked data for the 2010 season to estimate
lineman performance for different team-positions instead of individual
defenders (e.g., Chicago Bears' left tackle, Pittsburgh Steelers'
center, etc.). This work uses survival analysis to model time in pocket
and completion percentage for quarterbacks before proposing a measure
for linemen's contribution to their team's passing in terms of yards
gained.

\citet{Wolfson2017Forecasting} comment on the two aforementioned
articles that ``{[}a{]}lthough these are exciting preliminary steps,
there is still a long way to go before we can provide a comprehensive
appraisal of the achievements of an individual lineman.'' The challenge
here is fundamental, since there were not enough public data at the time
to develop any meaningful metric for linemen in football, as also noted
by \citet{AlamarGould2008}. However, with the granularity of player
tracking data, we have access to data not only for the linemen but also
for every player on the field. This provides us with a great opportunity
to study and gain better insights into linemen performance in football.

\hypertarget{our-contribution}{%
\subsection{Our Contribution}\label{our-contribution}}

In this paper, we focus on the evaluation of defensive linemen in
football. We propose STRAIN, a metric for measuring pass rush
effectiveness, inspired by the concept of strain rate in materials
science. Our statistic gives a continuous measure of pressure for every
pass rusher on the football field over the course of an entire play.
This allows for the assessment of pass rush success even on plays that
do not result in an observed outcome like a sack, hit, or hurry. We view
this as a major step forward for accurately evaluating defensive linemen
performance. We also demonstrate that STRAIN is a stable quantity over
time and predictive of defensive pressure. Additionally, we consider a
multilevel model to estimate every pass rusher's contribution to the
average STRAIN in a play while controlling for player positions, team,
and various play-level information. We note that although our focus in
this paper is on pass rushers, our approach can be extended to the
evaluation of pass blockers in American football.

The remainder of this manuscript is outlined as follows. We first
describe the player tracking data provided by the Big Data Bowl 2023 in
Section \ref{sec:data}. We then introduce the mathematical motivation
and definition of our measure STRAIN, followed by our modeling approach
in Section \ref{sec:methods}. Next, we present applications of STRAIN
and study different statistical properties of the metric in Section
\ref{sec:results}. We close with our discussion of future directions
related to this work in Section \ref{sec:discussion}.

\hypertarget{sec:data}{%
\section{Data}\label{sec:data}}

In the forthcoming analysis, we rely on the data from the NFL Big Data
Bowl 2023 provided by the NFL Next Gen Stats tracking system. The data
corresponds to 8,557 passing plays across 122 games in the first eight
weeks of the 2021 NFL regular season. For each play, we have information
on the on-field location, speed, angle, direction, and orientation of
each player on the field and the football at a rate of 10 Hz (i.e., 10
measurements per second), along with event annotations for each frame
such as ball snap, pass forward, and quarterback sack, to name a few.

For our investigation, we consider only the frames between the ball snap
and when a pass forward or quarterback sack is recorded for each play.
We also remove all plays with multiple quarterbacks on the field, since
we need a uniquely defined quarterback to compute our metric. After
preprocessing, there are 251,060 unique frames corresponding to moments
of time from the start of the play at snap until the moment the
quarterback either throws the pass or is sacked.

\begin{figure}

{\centering \includegraphics{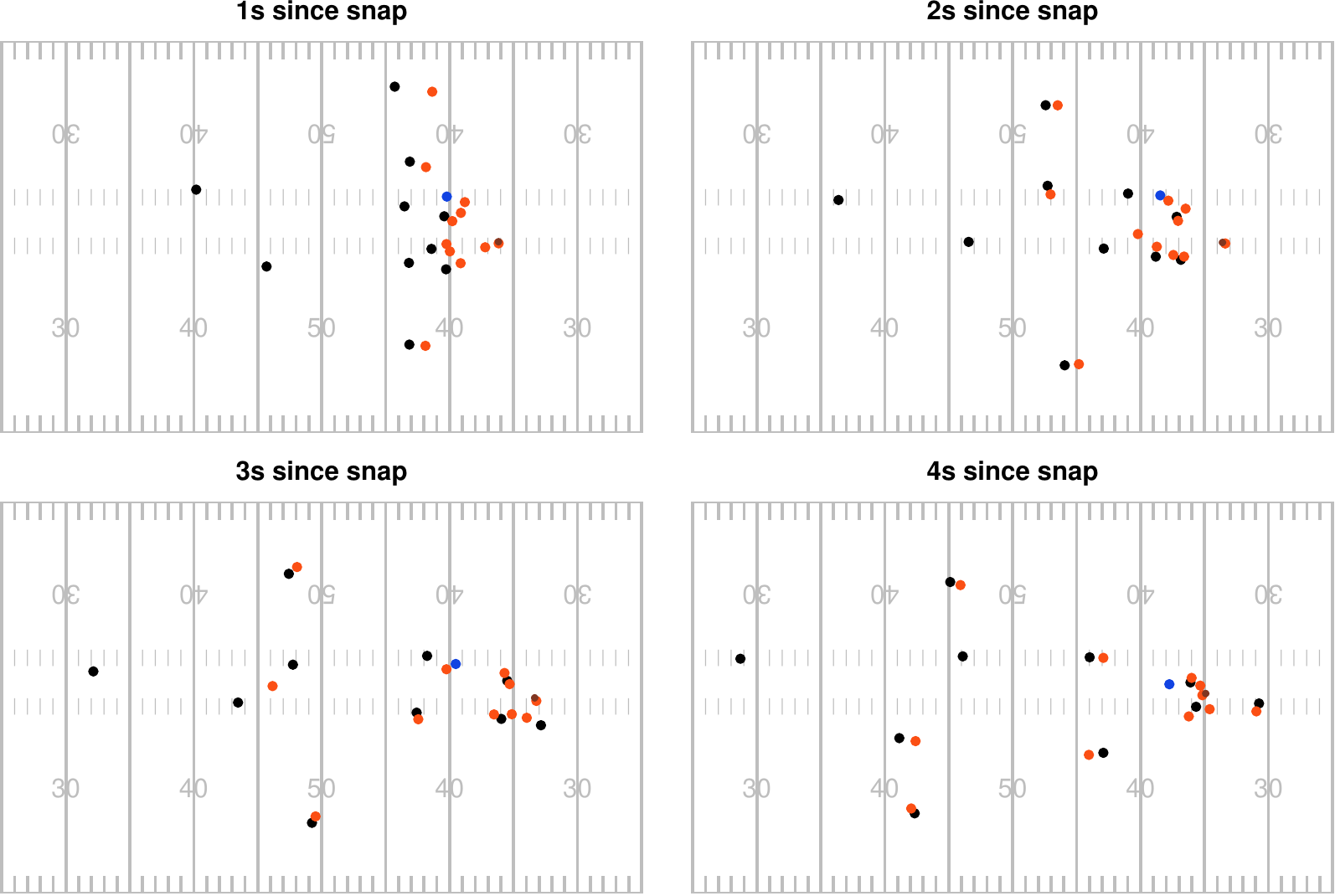} 

}

\caption{A display of the player tracking data for a play during the Las Vegas Raiders (defense, in black) versus Denver Broncos (offense, in orange) NFL game on October 17, 2021. Raiders defensive end Maxx Crosby is highlighted in blue. Snapshots are captured at 1, 2, 3, and 4 seconds after the ball snap.}\label{fig:fig_field_raw}
\end{figure}

Table \ref{tab:track} displays a tracking data example for a
\href{https://www.raiders.com/video/de-maxx-crosby-sacks-qb-teddy-bridgewater-for-a-loss-of-6-yards-nfl}{play}
from the 2021 NFL regular season week six matchup between the Las Vegas
Raiders and Denver Broncos, which ends with Broncos quarterback Teddy
Bridgewater getting sacked by Raiders defensive end Maxx Crosby. In
addition, Figure \ref{fig:fig_field_raw} presents the locations of every
Las Vegas (in black) and Denver (in orange) player on the field from
this play at 1, 2, 3 and 4 seconds after the ball snap, with Maxx Crosby
highlighted in blue.

\begin{table}
\caption{Example of tracking data for a play during the Las Vegas Raiders versus Denver Broncos NFL game on October 17, 2021. The data shown here are for Raiders defensive end Maxx Crosby, and the frames included are between the ball snap and when the sack by Crosby is recorded. \label{tab:track}}
\centering
\begin{tabular}{ccccccccc}
\hline
frameId & x & y & s & a & dis & o & dir & event \\ 
\hline
7 & 67.68 & 29.89 & 0.34 & 1.57 & 0.04 & 124.86 & 88.21 & ball\_snap \\ 
8 & 67.76 & 29.89 & 0.69 & 2.13 & 0.08 & 124.07 & 89.59 & None \\ 
\vdots & \vdots & \vdots & \vdots & \vdots & \vdots & \vdots & \vdots & \vdots \\ 
50 & 73.67 & 25.06 & 4.19 & 2.62 & 0.42 & 134.21 & 125.26 & qb\_sack \\ 
\hline
\end{tabular}
\end{table}

\begin{figure}

{\centering \includegraphics{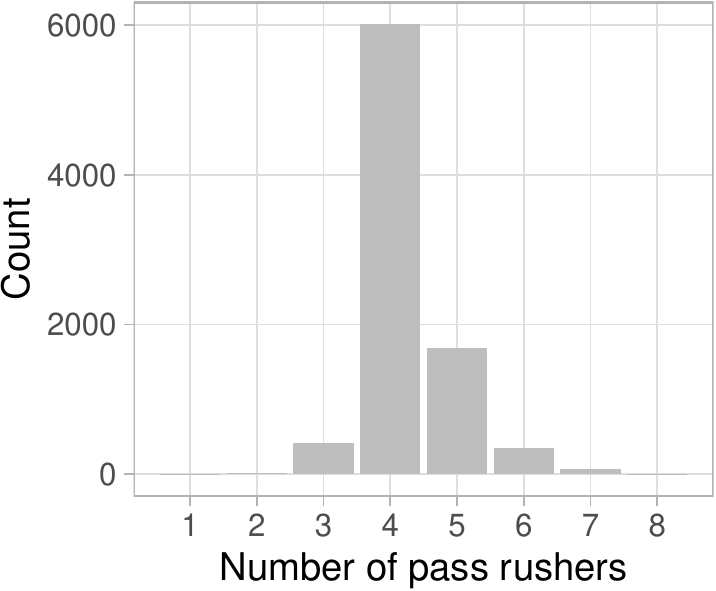} 

}

\caption{Distribution of the number of pass rushers on passing plays.}\label{fig:pass_rushers_count}
\end{figure}

Along with the tracking information, the Big Data Bowl 2023 includes
scouting data provided by Pro Football Focus (PFF). This contains
manually-collected player-level information, such as the player's role
(e.g., whether they are a pass rusher and pass blocker) and credited
events (e.g., player is credited with hitting the QB on the play). In
this manuscript, we use the PFF data to identify 36,362 unique pass rush
attempts by players designated in the ``pass rush'' role across all
plays. For context, Figure \ref{fig:pass_rushers_count} displays the
distribution of the number of observed pass rushers involved in a play,
ranging from 1 to 8 with 4 pass rushers (i.e., 4-man rush formation) as
the most common value. We also leverage this scouting data to count how
many hits, hurries, and sacks each pass rusher is credited with across
the span of observed data. Additionally, we use the PFF player roles to
identify the blocking matchup for each pass rusher in order to adjust
for opponent strength, as discussed in Section \ref{sec:methods}.

\hypertarget{sec:methods}{%
\section{Methods}\label{sec:methods}}

\hypertarget{sec:defn}{%
\subsection{Motivation and Definition of STRAIN}\label{sec:defn}}

In materials science, strain \citep{callister2018materials} is the
deformation of a material from stress, showing the change in a
material's length relative to its original length. Formally, let
\(L(t)\) be the distance between any given two points of interest within
a material at time \(t\), and \(L_0\) be the initial distance between
those two points. The strain for a material at time \(t\) is defined as
\begin{equation*}
\varepsilon(t)= \frac{L(t) - L_{0}}{L_{0}}.
\end{equation*} Notice that this measure is unitless due to being a
ratio of two quantities having the same unit.

Accordingly, the strain rate of a material measures the change in its
deformation with respect to time. Mathematically, the strain rate of a
material can be expressed as the derivative of its strain. That is,
\begin{equation*}
\varepsilon'(t) = \frac{d \varepsilon}{dt} = \frac{v(t)}{L_{0}},
\end{equation*} where \(v(t)\) is the velocity at which the two points
of interest within the material are moving away from or toward each
other. Whereas strain has no units, the strain rate is measured in
inverse of time, usually inverse second.

Motivated by its scientific definition, we draw a delightful analogy
between strain rate and pass rushing in football. Just as strain rate is
a measure of deformation in materials science, a pass rusher's efforts
involve the application of deformation against the offensive line, with
the ultimate goal of breaking through the protection to reach the
quarterback. The players can be viewed as ``particles'' in some material
and the defensive ``particles'' are attempting to exert pressure on the
pocket with the aim of compressing and collapsing this pocket around the
quarterback.

In order to apply strain rate to measure NFL pass rusher effectiveness,
we make modifications to how this concept is traditionally defined. Let
\((x_{ijt}, y_{ijt})\) be the \((x, y)\) location on the field of pass
rusher \(j = 1, \cdots, J\) at frame \(t = 1, \cdots, T_i\) for play
\(i = 1, \cdots, n\); and \((x^{QB}_{it}, y^{QB}_{it})\) be the
\((x, y)\) location of the quarterback at frame \(t\) during play \(i\).

\begin{itemize}
\tightlist
\item
  The distance between pass rusher \(j\) and the quarterback at frame
  \(t\) during play \(i\) is \begin{equation*}
  s_{ij}(t) = \sqrt{(x_{ijt} - x^{QB}_{it})^2 + (y_{ijt} - y^{QB}_{it})^2}.
  \end{equation*}
\item
  The velocity at which pass rusher \(j\) is moving toward the
  quarterback at frame \(t\) during play \(i\) is \begin{equation*}
  v_{ij}(t) = s'_{ij}(t) = \frac{ds_{ij}(t)}{dt}.
  \end{equation*}
\item
  The STRAIN for pass rusher \(j\) at frame \(t\) during play \(i\) is
  \begin{equation*}
  \text{STRAIN}_{ij}(t) = \frac{- v_{ij}(t)}{s_{ij}(t)}.
  \end{equation*}
\end{itemize}

Note that to distinguish our metric from strain and strain rate in
materials science, we write it in capital letters (STRAIN) for the
remainder of this manuscript.

Recall that based on its materials science property, an increase in
strain rate is associated with an increase in the distance between two
points. In the American football setting, the two points of interest are
the pass rusher and the quarterback, and we expect our metric to
increase as the distance between the pass rusher and the quarterback
decreases. Thus, the negative sign in the numerator of our formula
effectively accounts for this. Additionally, rather than keeping the
initial distance (\(L_0\) as previously denoted) between two points
constant over time, we update the initial position to be the player
locations at the beginning of each frame. This gives us the STRAIN for
each frame throughout a play.

Since we only observe the distance and velocity quantities discretely in
increments of 10 frames/second, a point estimate for our proposed metric
STRAIN for pass rusher \(j\) at frame \(t\) during play \(i\) is
\begin{equation*}
\widehat{\text{STRAIN}}_{ij}(t) = \cfrac{-\cfrac{s_{ij}(t) - s_{ij}(t - 1)}{0.1}}{s_{ij}(t)}.
\end{equation*} Notice that this quantity increases in two ways: 1) the
rate at which the rusher is moving toward the quarterback increases,
and 2) the distance between the rusher and the quarterback decreases.
Both of these are indications of an effective pass rush attempt.
Finally, our statistic STRAIN is measured in inverse second, similar to
strain rate. Note that the reciprocal of our metric (1/STRAIN) has an
interesting and straightforward interpretation: the amount of time
required for the rusher to get to the quarterback at the current
location and rate at any given time \(t\).

Moreover, since we observe STRAIN at every tenth of a second within each
play, we can then compute the average STRAIN across all frames played
for every pass rusher. Formally, the average STRAIN, denoted by
\(\overline{\text{STRAIN}}\), for pass rusher \(j\) involved in \(n_j\)
total plays across \(\sum_{i\in Z_j} T_i\) total frames, where \(Z_j\)
is the set of all plays with pass rusher \(j\)'s involvement, is
\begin{equation*}
\overline{\text{STRAIN}}_{j} = \frac{1}{\sum_{i\in Z_j} T_i} \sum_{i\in Z_j} \sum_{t=1}^{T_i} \widehat{\text{STRAIN}}_{ij}(t).
\end{equation*} This can be helpful for player evaluation, as we
determine the most effective pass rushers based on their average STRAIN
values in Section \ref{sec:strainbar}. We also use average STRAIN to
assess different statistical properties of our metric in Section
\ref{sec:statprop}.

\hypertarget{sec:multilevel}{%
\subsection{Multilevel Model for Play-Level
STRAIN}\label{sec:multilevel}}

In addition to the average STRAIN over all frames played, we can also
calculate pass rusher \(j\)'s observed average STRAIN on a single play
\(i\) consisting of \(T_i\) total frames, \begin{equation*}
\overline{\text{STRAIN}}_{ij} = \frac{1}{T_i} \sum_{t=1}^{T_i} \widehat{\text{STRAIN}}_{ij}(t).
\end{equation*} While this aggregated measure is a simple first step for
pass rush evaluation, the observed average STRAIN on a single play is
likely due to numerous factors. Besides the pass rusher's ability, there
is variability in the opposing strength of pass blockers across plays a
pass rusher is involved, both at the individual and team levels. Thus,
we need to appropriately divide the credit of an observed average STRAIN
across the different players and team involved, amongst other factors.

To this end, we fit a multilevel model to evaluate pass rushers' impact
on the average STRAIN observed in a play, while accounting for their
team on defense, the opposing team on offense, and their assigned pass
blocker. We identify the pass blocker linked with the pass rusher of
interest using the scouting data provided by PFF as mentioned in Section
\ref{sec:data}. Since there can be multiple blockers matching up with a
rusher, for simplicity, we consider the nearest blocker positioned to
the pass rusher at the start of the play. We use random intercepts for
the two player groups: pass rushers as \(R\) and nearest pass blockers
as \(B\), as well as for the two team groups: defense \(D\) and offense
\(O\). We also account for attributes about pass rusher \(j\) in play
\(i\) through the covariate vector \(\mathbf{x_{ij}}\), and estimate
their respective coefficients \(\boldsymbol \beta\) as fixed effects.
Our model for the average STRAIN by pass rusher \(j\) on play \(i\) is
as follows. \begin{equation*}
\begin{aligned}
\overline{\text{STRAIN}}_{ij} &\sim N(R_{j[i]} + B_{b[ij]} + D_{d[i]} + O_{o[i]} + \mathbf{x_{ij}} \boldsymbol{\beta}, \sigma^2), \text{ for } i = 1, \dots, n \text{ plays} \\
R_{j} &\sim N(\mu_R, \sigma^2_R), \text{ for } j = 1, \cdots, \text{ \# of pass rushers}, \\
B_{b} &\sim N(\mu_B, \sigma^2_B), \text{ for } b = 1, \cdots, \text{ \# of pass blockers}, \\
D_{d} &\sim N(\mu_D, \sigma^2_D), \text{ for } d = 1, \cdots, \text{ \# of defensive teams}, \\
O_{o} &\sim N(\mu_O, \sigma^2_O), \text{ for } o = 1, \cdots, \text{ \# of offensive teams}.
\end{aligned}
\end{equation*}

In detail, we consider a normal distribution to shrink the random
intercepts for each player and team toward their respective group means.
This is a useful property since we do not observe the same number of
plays for each player. For the team effects, this provides us with the
average defense and offense team-level effects on a pass rusher's
STRAIN. Due to the nested nature of players on teams, the individual
pass rusher and blocker random intercepts reflect the respective
player's effect relative to their team effects. We implement the model
using penalized likelihood via the \texttt{lme4} package in \texttt{R}
\citep{bateslme4, R2023Language}.

In order to provide a measure of uncertainty for our random effects, we
use a bootstrapping strategy similar to the approach in
\citet{yurko2019nflwar}. Specifically, we resample team drives within
games, which preserves the fact that team schedules are fixed but allows
for random variation in player usage since this is dependent on team
decision making. By resampling plays within the same drive together,
this allows us to generate realistic simulated data in comparison to
sampling individual plays. For each bootstrapped dataset, we fit the
aforementioned multilevel model to obtain a distribution of estimates
for the considered player and team effects.

As for the fixed effects about pass rusher \(j\) in play \(i\), we
include a variety of features that likely contribute to variation in
STRAIN. First, we adjust for the position of both the pass rusher and
nearest blocker to account for any positional effects. Table
\ref{tab:pos} shows our positional categorization for the pass rush and
pass block roles. These are encoded as indicator variables with
defensive ends and tackles as the reference levels for the pass rushers
and blockers, respectively. We also account for the number of pass
blockers on the play, since teams may decide to employ a more protective
scheme that could lower the observed STRAIN. Finally, we control for
play-context covariates with respect to the offensive team. These
include the current down (first, second, third, fourth, or two-point
conversion), yards to go for a first down, and current yardline (i.e.,
distance from the possession team's goal line). We consider play-context
information since these variables impact a team's designed play, which
may result in a play with low or high STRAIN regardless of the pass
rusher's role. For instance, a team may call a short pass that is
intended to be thrown early which could limit the amount of STRAIN on a
play. Or a team may need to throw a deep pass which would require more
time and potentially create more STRAIN. We do not account for time
directly in the model due to the concern that the time it takes for a
quarterback to throw the ball is itself a function of both the play call
and pressure from pass rushers. Thus, since we do not know the designed
play call, we condition on the play context to adjust for play-level
differences attributing to a pass rusher's STRAIN.

\begin{table}
\caption{Position groupings for pass rushers and blockers. \label{tab:pos}}
\centering
\begin{tabular}{ll}
\hline
Role & Position \\ 
\hline
Pass rush & Defensive end \\ 
Pass rush & Defensive tackle \\ 
Pass rush & Nose tackle \\ 
Pass rush & Outside linebacker \\ 
Pass rush & Interior linebacker (middle linebacker, inside linebacker) \\ 
Pass rush & Secondary (cornerback, free safety, strong safety) \\
\hline
Pass block & Center \\ 
Pass block & Guard \\ 
Pass block & Tackle \\ 
Pass block & Other (tight end, running back, fullback, wide receiver) \\ 
\hline
\end{tabular}
\end{table}

\hypertarget{sec:results}{%
\section{Results}\label{sec:results}}

\hypertarget{real-game-illustration-of-strain}{%
\subsection{Real-Game Illustration of
STRAIN}\label{real-game-illustration-of-strain}}

\begin{figure}

{\centering \includegraphics{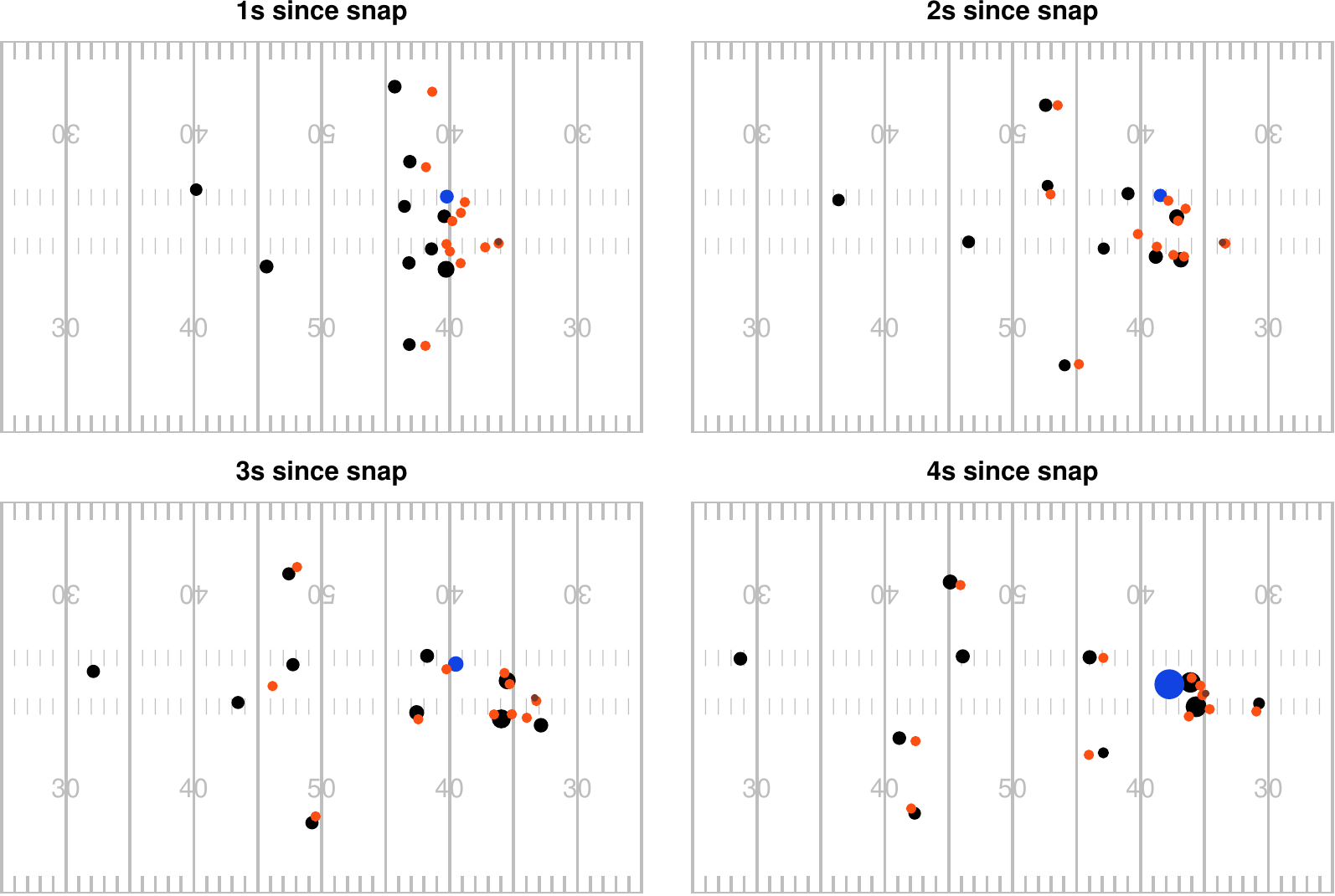} 

}

\caption{A display of the player tracking data for a play during the Las Vegas Raiders (defense, in black) versus Denver Broncos (offense, in orange) NFL game on October 17, 2021. Raiders DE Maxx Crosby is highlighted in blue. For each Raiders defender, the point size indicates their individual STRAIN value, with larger points suggesting larger STRAIN. Snapshots are captured at 1, 2, 3, and 4 seconds after the ball snap.}\label{fig:fig_field}
\end{figure}

\begin{figure}

{\centering \includegraphics{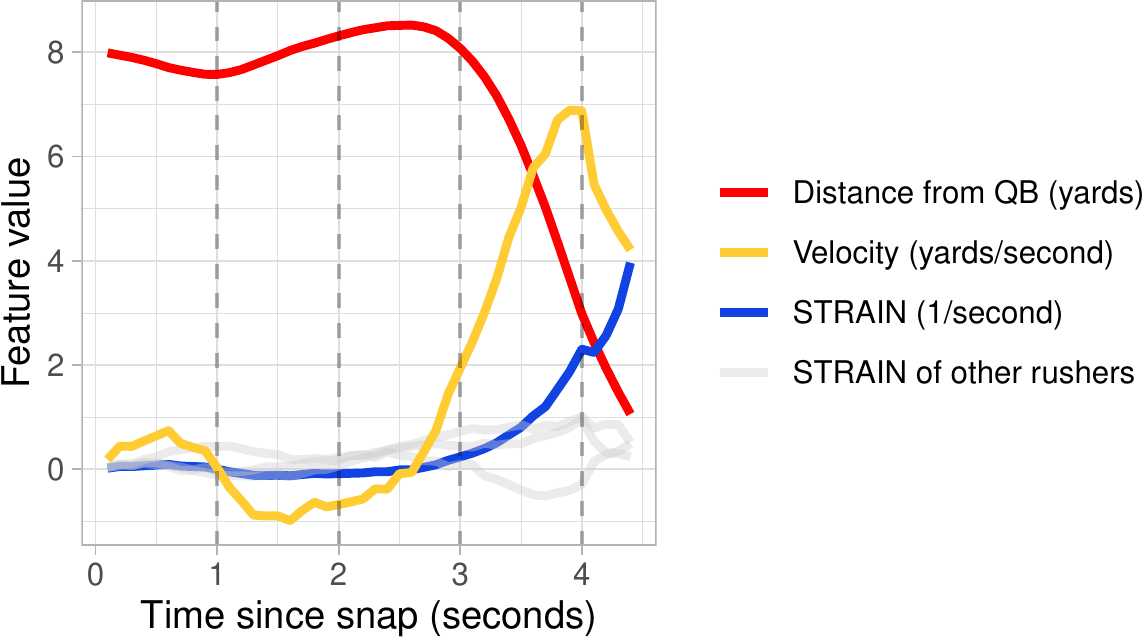} 

}

\caption{Changes in STRAIN, velocity, and distance from the quarterback for Maxx Crosby over the course of a successful pass rush play that results in a sack. The STRAIN for other pass rushers throughout this play is also displayed.}\label{fig:fig_crosby_curves}
\end{figure}

To illustrate our proposed metric STRAIN for pass rush evaluation, we
use the same play from the Las Vegas Raiders and Denver Broncos game as
mentioned in Section \ref{sec:data}. Figure \ref{fig:fig_field} shows an
updated version of Figure \ref{fig:fig_field_raw}, with the point size
for each Las Vegas defender corresponding to the estimated STRAIN in the
selected frames as the play progresses. This is accompanied by Figure
\ref{fig:fig_crosby_curves}, which is a line graph showing how Crosby's
distance from the quarterback, velocity, and STRAIN change continuously
throughout the play.

We observe that for the first two seconds, Crosby is being blocked by a
Denver offensive lineman and unable to get close to the quarterback,
hence the corresponding STRAIN values are virtually zero. Suddenly at
around three seconds after the snap, the STRAIN for Crosby starts to
increase after the Raiders defensive end is freed up and charges toward
Bridgewater. At 4 seconds after the snap, Crosby's STRAIN is 2.30, which
means at his current moving rate, it will take Crosby about 0.43
(1/2.30) seconds to make the distance between him and the quarterback 0
(i.e.~essentially sack the quarterback). This matches well with the
final outcome of the play, as the sack takes place at the very last
frame (4.4 seconds) where the estimated STRAIN for Crosby reaches its
peak at 3.96.

Moreover, Figure \ref{fig:fig_crosby_curves} clearly demonstrates the
interactions between the features for Maxx Crosby. Here, a higher STRAIN
generally corresponds to faster moving rate toward the quarterback.
STRAIN also increases as the distance between the pass rusher and
quarterback is being reduced. Both of these relationships suggest an
overall successful pass rush by Maxx Crosby. It is also notable that
Crosby, who is credited with a sack, generates more STRAIN than other
pass rushers during this play.

\begin{figure}

{\centering \includegraphics{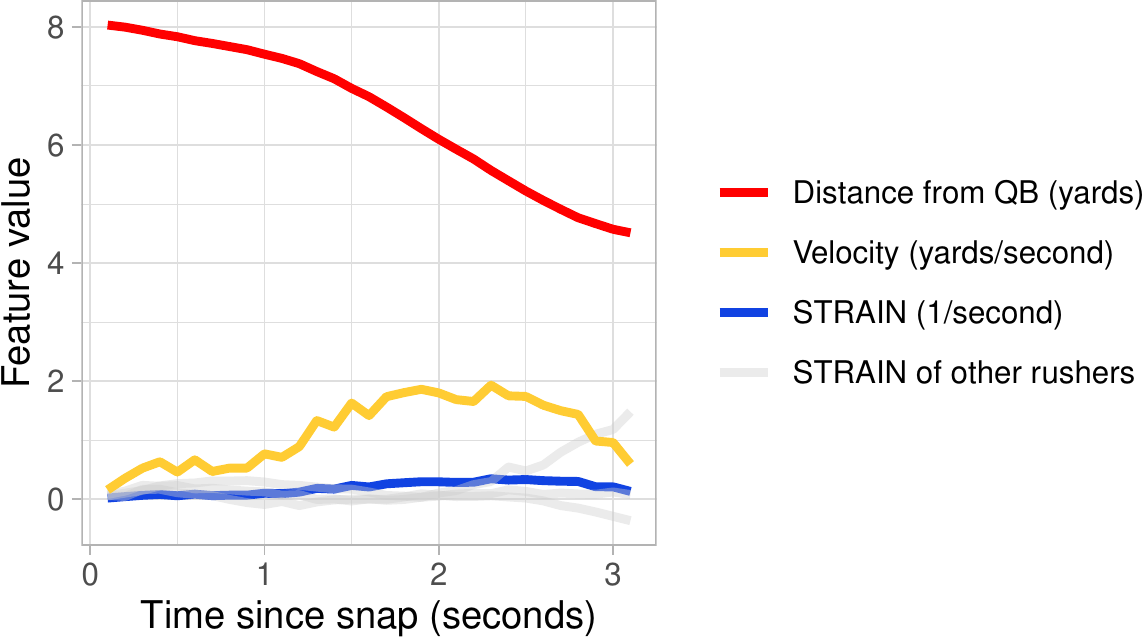} 

}

\caption{Changes in STRAIN, velocity, and distance from the quarterback for Maxx Crosby over the course of an unsuccessful pass rush play. The STRAIN for other pass rushers throughout this play is also displayed.}\label{fig:fig_crosby_curves_unsucess}
\end{figure}

In contrast, Figure \ref{fig:fig_crosby_curves_unsucess} shows the
feature curves for an unsuccessful pass rush attempt by the Raiders
defense from the same game. In this
\href{https://www.nfl.com/videos/teddy-bridgewater-courtland-sutton-barely-miss-out-on-chance-at-62-yard-td-bomb}{play},
Broncos QB Teddy Bridgewater is well-protected by the offensive line and
is able to release a long pass to a receiver. We see that the STRAIN
generated by Crosby is relatively small over the course of this play,
compared to the previous play (as shown in Figure
\ref{fig:fig_crosby_curves}) which results in a sack.

\hypertarget{sec:positional}{%
\subsection{Positional STRAIN Curves}\label{sec:positional}}

\begin{figure}

{\centering \includegraphics{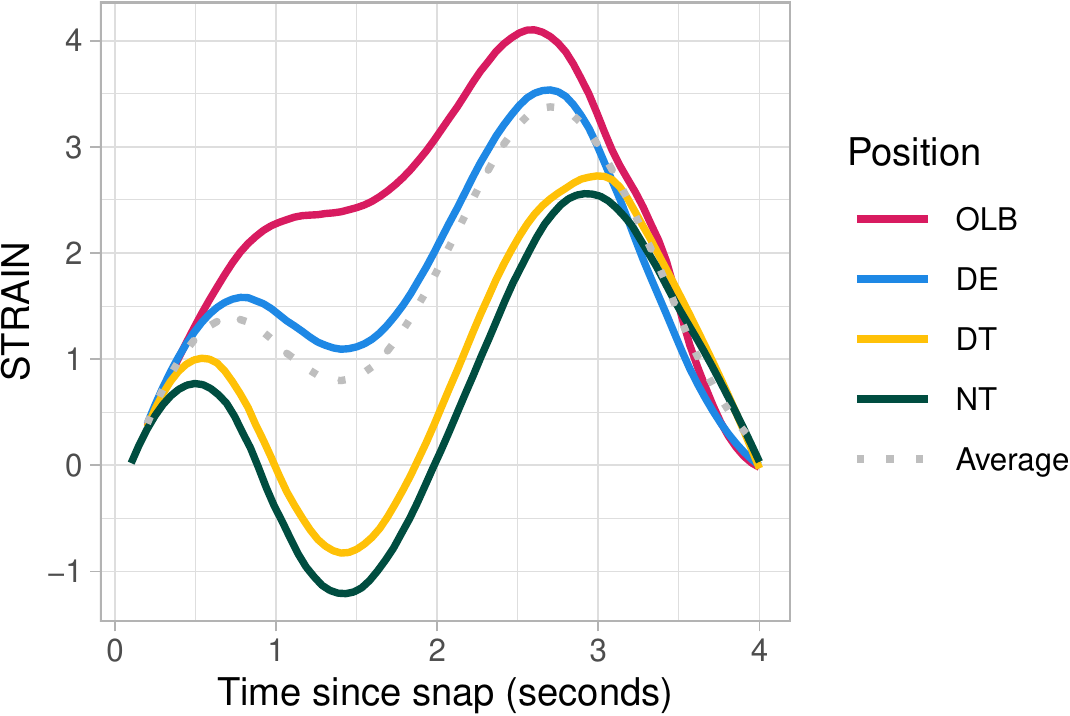} 

}

\caption{STRAIN curves for different positions. Edge rushers (outside linebackers and defensive ends) tend to generate more STRAIN than interior rushers (defensive tackles and nose tackles). The dotted gray line represents the average STRAIN curve for all players without accounting for position.}\label{fig:fig_pos_curves}
\end{figure}

\begin{figure}

{\centering \includegraphics{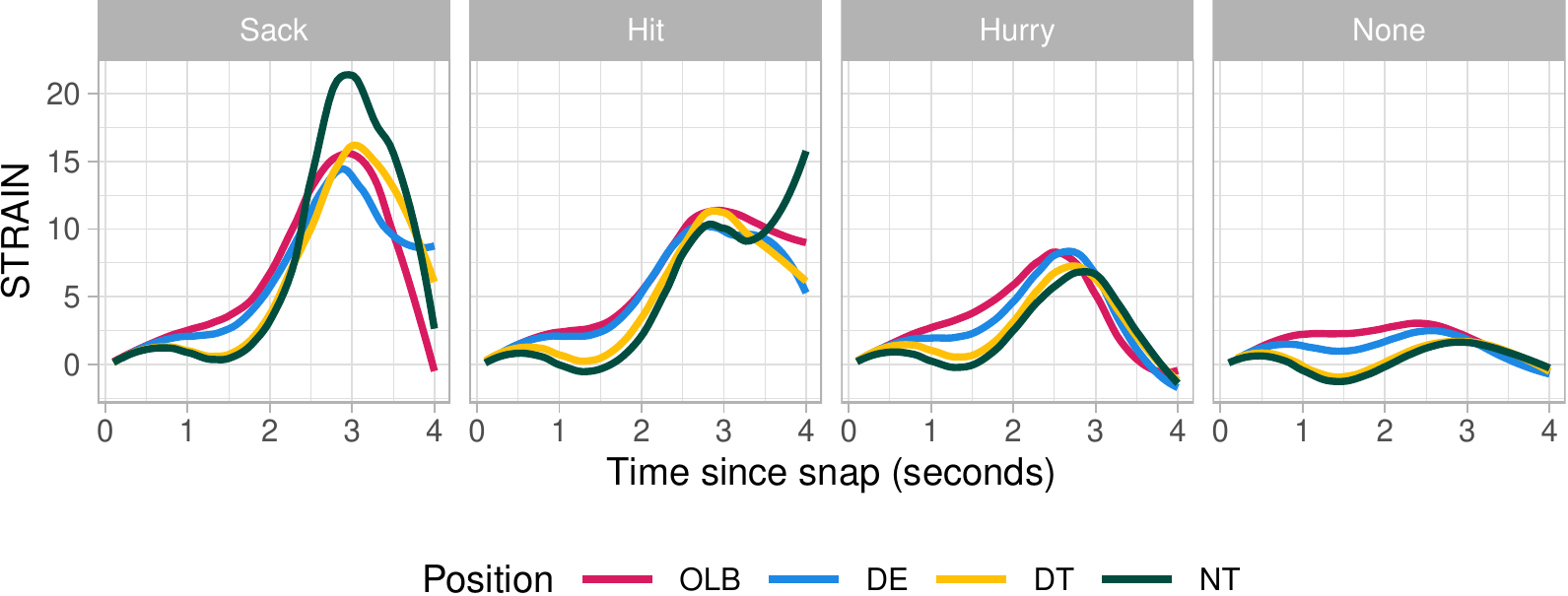} 

}

\caption{Positional STRAIN curves by play outcome (sack, hit, hurry, and none).}\label{fig:fig_outcome_curves}
\end{figure}

Figure \ref{fig:fig_pos_curves} displays the average STRAIN by position
for the first 40 frames (4 seconds) after the snap. For each position,
the curve is based on the average of the observed STRAIN at each frame
across all plays and players within the position. We observe a clear
difference in STRAIN between edge rushers (outside linebackers and
defensive ends) and interior linemen (defensive tackles and nose
tackles). Specifically, edge rushers have higher STRAIN than interior
linemen on average, as they are more easily able to approach the
quarterback on the edge of the pocket versus, for instance, a nose
tackle attacking the line head on.

On average, STRAIN appears to increase for the first 0.5 seconds of a
play, followed by a decline in the next second. STRAIN then increases
again until around 2.5--3 seconds after the snap before trending down
toward the end of the play. In context, this reflects the actions that
a pass rusher initially moves toward the quarterback, but is then
stopped by the offensive line while the quarterback drops back. When the
quarterback stops dropping, the rusher closes the gap and increases
STRAIN, before slowing down later on.

Further, Figure \ref{fig:fig_outcome_curves} shows the positional STRAIN
curves by whether a play's outcome is a hit, sack, hurry, or none of
those. We clearly see that players tend to generate more STRAIN when a
play ends in a hit, sack, or hurry, compared to no outcome. Within the
three pressure metrics, it is not surprised that a sack corresponds to
the highest amount of STRAIN, followed by a hit and hurry.

\hypertarget{sec:strainbar}{%
\subsection{Ranking the Best Pass Rushers}\label{sec:strainbar}}

Since STRAIN is observed continuously for every play in our data, this
allows us to aggregate across all frames played and compute the average
STRAIN for NFL pass rushers over the course of the eight-week sample
size, as discussed in Section \ref{sec:defn}. Based on the clearly
distinct patterns for different positions as previously observed, we
evaluate interior pass rushers (nose tackles and defensive tackles)
separately from edge rushers (outside linebackers and defensive ends).
Tables \ref{tab:edge} and \ref{tab:interior} are leaderboards for the
NFL's best edge and interior rushers (with at least 100 plays) rated by
the average STRAIN across all frames for the first eight weeks of the
2021 regular season. The tables also consist of the total number of
hits, hurries, and sacks (determined from PFF scouting data) for each
defender.

Our results are mostly consistent with conventional rankings of rushers.
Notably, Myles Garrett and TJ Watt are widely recognized as top-tier
edge rushers and both show up in our top edge rusher list; whereas Aaron
Donald, who is undoubtedly the best interior defender in football,
appears at the top of our interior rusher rankings. Moreover, our
leaderboards largely match the rankings of experts in the field. For
instance, there is considerable overlap between our lists and PFF's edge
\citep{Monson2022nfl} and interior \citep{Linsey2022nfl} rusher rankings
released after the 2021 season. This ultimately lends credibility to our
proposed metric STRAIN as a measure of pass rushing effectiveness.

\begin{table}
\caption{Top 15 edge rushers (with at least 100 snaps played) according to the average STRAIN across all frames played. \label{tab:edge}}
\centering
\begin{tabular}{rlllrrrrr}
\hline
Rank & Player & Team & Position & Snaps & Hits & Hurries & Sacks & $\overline{\text{STRAIN}}$ \\
\hline
1 & Rashan Gary & GB & OLB & 176 & 10 & 25 & 5 & 2.82 \\ 
2 & Leonard Floyd & LA & OLB & 185 & 2 & 25 & 8 & 2.80 \\ 
3 & Justin Houston & BAL & OLB & 132 & 8 & 8 & 4 & 2.78 \\ 
4 & Myles Garrett & CLE & DE & 197 & 9 & 29 & 12 & 2.75 \\ 
5 & Von Miller & DEN & OLB & 145 & 4 & 21 & 5 & 2.75 \\ 
6 & T.J. Watt & PIT & OLB & 147 & 6 & 9 & 8 & 2.71 \\ 
7 & Yannick Ngakoue & LV & DE & 175 & 6 & 20 & 4 & 2.70 \\ 
8 & Alex Highsmith & PIT & OLB & 129 & 4 & 7 & 2 & 2.65 \\ 
9 & Preston Smith & GB & OLB & 124 & 4 & 8 & 2 & 2.61 \\ 
10 & Randy Gregory & DAL & DE & 134 & 7 & 19 & 5 & 2.58 \\ 
11 & Joey Bosa & LAC & OLB & 160 & 5 & 21 & 4 & 2.58 \\ 
12 & Darrell Taylor & SEA & DE & 107 & 5 & 9 & 3 & 2.57 \\ 
13 & Josh Sweat & PHI & DE & 159 & 4 & 14 & 5 & 2.57 \\ 
14 & Maxx Crosby & LV & DE & 198 & 12 & 30 & 7 & 2.56 \\ 
15 & Markus Golden & ARI & OLB & 164 & 5 & 14 & 5 & 2.50 \\ 
\hline
\end{tabular}
\end{table}

\begin{table}
\caption{Top 15 interior rushers (with at least 100 snaps played) according to the average STRAIN across all frames played. \label{tab:interior}}
\centering
\begin{tabular}{rlllrrrrr}
\hline
Rank & Player & Team & Position & Snaps & Hits & Hurries & Sacks & $\overline{\text{STRAIN}}$ \\
\hline
1 & Aaron Donald & LA & DT & 239 & 8 & 24 & 6 & 1.67 \\ 
2 & Solomon Thomas & LV & DT & 115 & 7 & 11 & 3 & 1.51 \\ 
3 & Quinton Jefferson & LV & DT & 144 & 6 & 8 & 3 & 1.46 \\ 
4 & Chris Jones & KC & DT & 139 & 3 & 18 & 3 & 1.42 \\ 
5 & DeForest Buckner & IND & DT & 198 & 4 & 18 & 4 & 1.26 \\ 
6 & Cameron Heyward & PIT & DT & 188 & 2 & 22 & 3 & 1.25 \\ 
7 & Javon Hargrave & PHI & DT & 156 & 6 & 15 & 5 & 1.24 \\ 
8 & Jerry Tillery & LAC & DT & 171 & 4 & 7 & 3 & 1.16 \\ 
9 & Ed Oliver & BUF & DT & 133 & 4 & 12 & 1 & 1.15 \\ 
10 & Osa Odighizuwa & DAL & DT & 162 & 3 & 18 & 3 & 1.13 \\ 
11 & Greg Gaines & LA & NT & 111 & 2 & 13 & 2 & 1.11 \\ 
12 & Leonard Williams & NYG & DT & 226 & 4 & 14 & 6 & 1.03 \\ 
13 & Christian Barmore & NE & DT & 166 & 5 & 17 & 1 & 1.02 \\ 
14 & Vita Vea & TB & NT & 184 & 6 & 12 & 1 & 1.01 \\ 
15 & B.J. Hill & CIN & DT & 123 & 3 & 4 & 3 & 0.96 \\ 
\hline
\end{tabular}
\end{table}

\hypertarget{sec:statprop}{%
\subsection{Statistical Properties of STRAIN}\label{sec:statprop}}

Next, we examine different statistical properties of our proposed metric
STRAIN. Our focus here is to understand the stability and predictability
of STRAIN, and we pose the following question: How much does our metric
vary from week to week? In other words, is previous performance
predictive of future performance based on our metric? Below, we attempt
the answer these questions, and the following results are for pass
rushers with at least 100 snaps played during the first eight weeks of
the 2021 NFL regular season.

\begin{figure}

{\centering \includegraphics{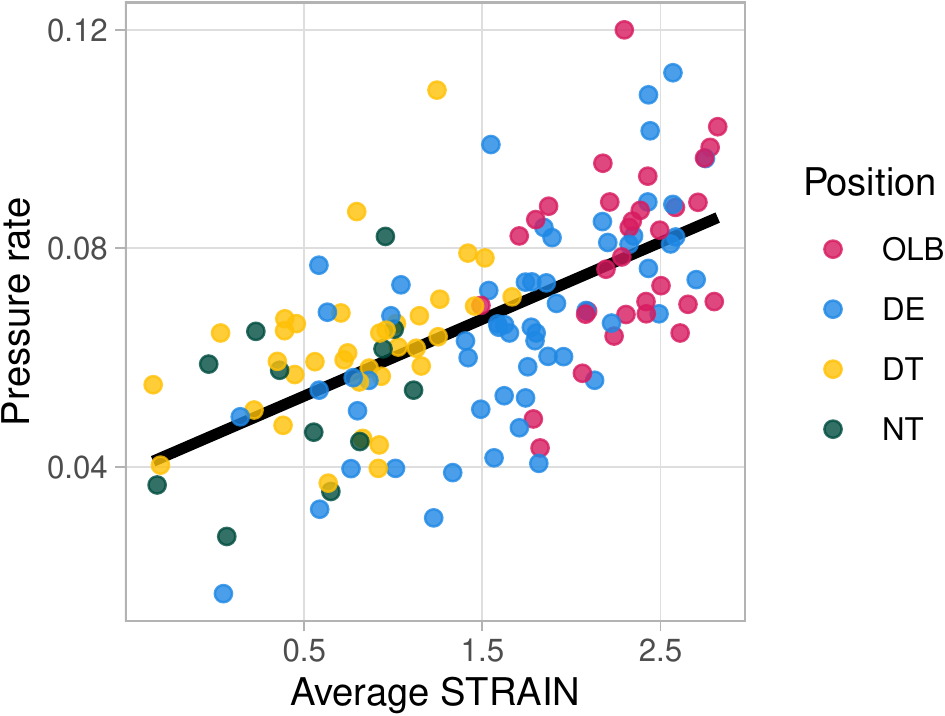} 

}

\caption{Relationship between average STRAIN and pressure rate (total hits, sacks, and hurries per snap) over the first eight weeks of the 2021 NFL season. There is a moderately strong association between average STRAIN and pressure rate ($r =0.6255$). Results shown here are for pass rushers with at least 100 snaps played over the eight-week data.}\label{fig:fig_cor_pressure}
\end{figure}

\begin{figure}

{\centering \includegraphics{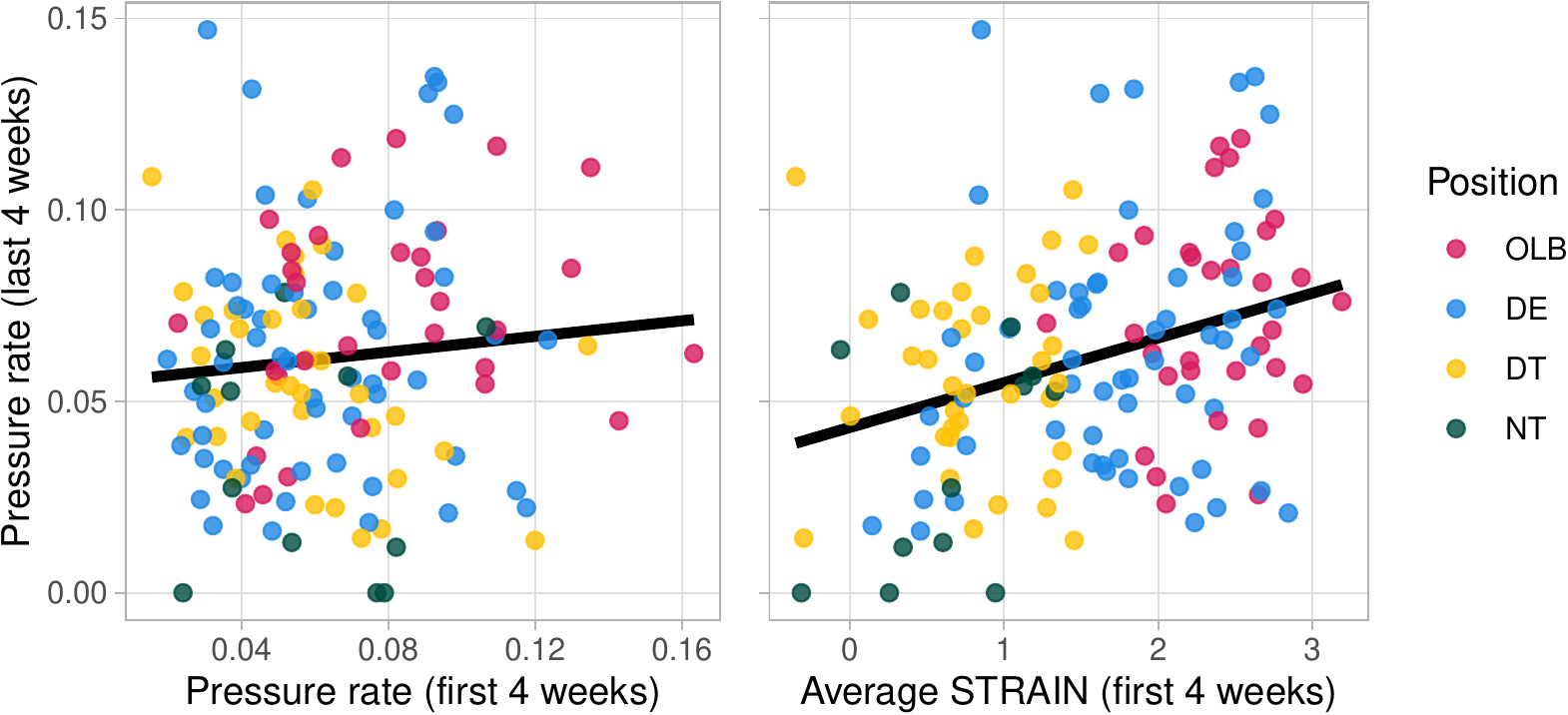} 

}

\caption{Relationships between pressure rate for the last 4 weeks of the 2021 NFL season and first four weeks' pressure rate (left) and average STRAIN (right). STRAIN is more predictive ($r=0.3217$) of future pressure rate than previous pressure rate itself ($r=0.0965$). Results shown here are for pass rushers with at least 100 snaps played over the eight-week data.}\label{fig:fig_predictability}
\end{figure}

We first investigate the predictability of STRAIN as a measure of
pressure. In particular, we look at how well our metric correlates with
a simple measure of pressure rate, defined as the total hits, sacks and
hurries per snap. Figure \ref{fig:fig_cor_pressure} is a scatterplot of
average STRAIN and pressure rate over the course the provided eight-game
sample, which reveals a fairly strong correlation (\(r =0.6255\))
between the quantities. Hence, defenders with high STRAIN values also
tend to generate more pressure toward the quarterback. Furthermore, the
average STRAIN for the first four weeks of the 2021 season is more
predictive of the last four weeks' pressure rate (\(r=0.3217\)) than the
first four weeks' pressure rate (\(r=0.0965\)), as illustrated in Figure
\ref{fig:fig_predictability}.

\begin{figure}

{\centering \includegraphics{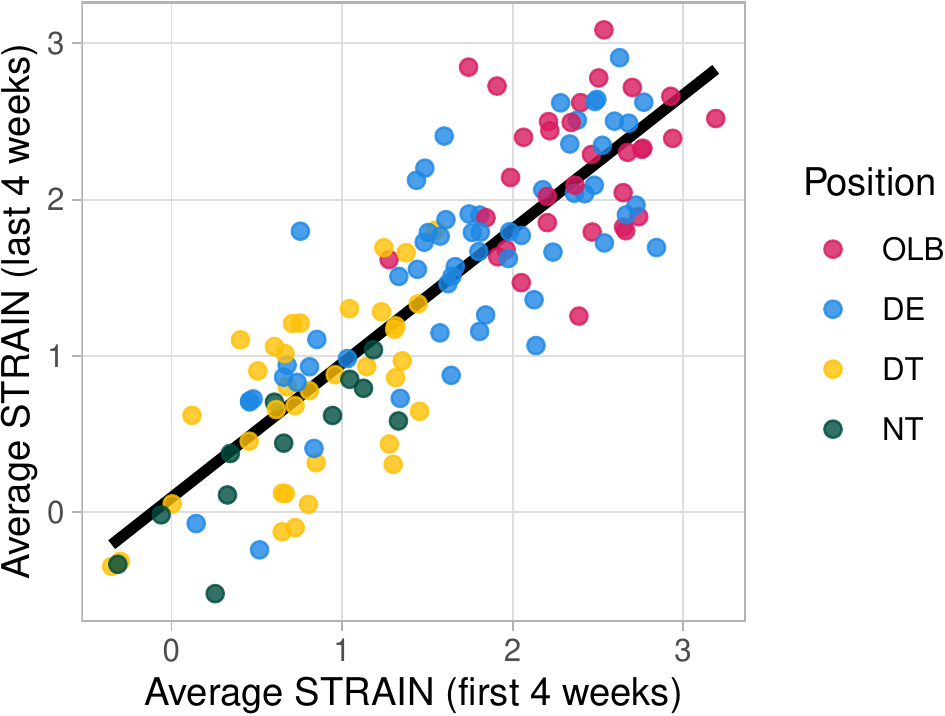} 

}

\caption{Relationship between average STRAIN for the last and first four weeks of the 2021 NFL season. A strong linear correlation ($r = 0.8545$) demonstrates that STRAIN is a highly stable metric over time. Results shown here are for pass rushers with at least 100 snaps played over the eight-week data.}\label{fig:fig_stability}
\end{figure}

We also analyze the stability of STRAIN over time by comparing the
average STRAIN across all frames played for the first and last four
weeks of the 2021 NFL regular season. It is apparent from Figure
\ref{fig:fig_stability} that there is a strong positive correlation
\((r = 0.8545)\) for this relationship. This means that STRAIN is a
highly stable football metric over the provided eight-week time window,
and pass rushers appear to carry their STRAIN values with them from week
to week. Overall, STRAIN performs well in both explaining defensive
pressure on the field and predicting future performance of pass rushers.

\hypertarget{multilevel-model-results}{%
\subsection{Multilevel Model Results}\label{multilevel-model-results}}

The results of fitting the multilevel model described in Section
\ref{sec:multilevel} is displayed in Table \ref{tab:fixedeff}. First, we
investigate the fixed effects terms of this model. It appears that that
the average STRAIN decreases by 0.7366 (95\% CI {[}-0.7875, -0.6858{]})
for every additional blocker in the offensive unit, after accounting for
other covariates. In other words, NFL pass rushers tend to generate more
pressure when facing fewer number of blockers, which makes intuitive
sense.

As for play context, defensive pressure appears to increase when more
yardage is required for the offense to reach a first down and for later
plays within a set of down, when all other predictors are held constant.
In particular, every extra yard in the distance needed for a first down
is associated with a 0.0491 increase (95\% CI {[}0.0426, 0.0555{]}) in
the average STRAIN. Relative to first down situations, higher amount of
pressure seem to happen in plays that come after (second, third, and
fourth downs). In addition, we have insufficient evidence for a
relationship between average STRAIN and the current yardline on the
field for the offensive team.

Moreover, we observe statistically significant differences among the
pass rush and pass block positions in most cases, while controlling for
other variables. The coefficient estimates for the pass rush position
terms reveal that players with positions closer to the line of scrimmage
(defensive tackles and nose tackles) tend to generate less STRAIN
(relative to defensive ends) than those lining up further back, which is
consistent with our results in Section \ref{sec:positional}. For
blockers, centers are those that absorb the most pressure on average,
more than guards and other offensive positions (compared to the baseline
level, tackles).

Next, Table \ref{tab:icc} displays the intraclass correlation
coefficients (ICC) for the four different groupings, describing the
proportion variance explained between the group terms in comparison to
the residual variance. While the residual variance is unsurprisingly the
largest value, between the player and team factors, we observe the
largest ICC for pass rushers, followed by the offensive and defensive
teams, and the blocker. This emphasizes how STRAIN is mostly
attributable to pass rushers, but is necessary to adjust for opposition
and other factors.

\begin{table}
\caption{Fixed effects coefficient estimates for the multilevel model for average STRAIN. Note that the reference down level is first down, the reference pass rusher level is defensive end (DE), denoted R:DE; the reference pass blocker level is tackle (T), denoted B:T. \label{tab:fixedeff}}
\centering
\begin{tabular}{lrrrr}
\hline
& estimate & se & $t$-statistic & $p$-value \\ 
\hline
Intercept & 1.8004 & 0.0991 & 18.1723 & 0.0000 \\ 
Number of blockers & -0.7366 & 0.0259 & -28.4153 & 0.0000 \\
Yards to go & 0.0491 & 0.0033 & 14.9013 & 0.0000 \\
Current yardline & 0.0001 & 0.0005 & 0.1844 & 0.8537 \\
$I_{\text{\{2nd down\}}}$ & 0.5143 & 0.0290 & 17.7439 & 0.0000 \\
$I_{\text{\{3rd down\}}}$ & 0.9176 & 0.0324 & 28.3613 & 0.0000 \\
$I_{\text{\{4th down\}}}$ & 0.4649 & 0.0767 & 6.0637 & 0.0000 \\
$I_{\text{\{2pt conversion\}}}$ & -0.1384 & 0.1797 & -0.7699 & 0.4413 \\
$I_{\text{\{R:DT\}}}$ & -0.7166 & 0.0678 & -10.5658 & 0.0000 \\
$I_{\text{\{R:interior\}}}$ & 0.5611 & 0.0999 & 5.6143 & 0.0000 \\
$I_{\text{\{R:NT\}}}$ & -0.9426 & 0.1021 & -9.2344 & 0.0000 \\
$I_{\text{\{R:OLB\}}}$ & 0.5574 & 0.0726 & 7.6742 & 0.0000 \\
$I_{\text{\{R:secondary\}}}$ & 1.4252 & 0.1114 & 12.7899 & 0.0000 \\
$I_{\text{\{B:C\}}}$ & -0.2983 & 0.0594 & -5.0237 & 0.0000 \\
$I_{\text{\{B:G\}}}$ & -0.1546 & 0.0471 & -3.2847 & 0.0000 \\ 
$I_{\text{\{B:other\}}}$ & -0.1260 & 0.0646 & -1.9508 & 0.0514 \\
\hline
\end{tabular}
\end{table}

\begin{table}
\caption{Intraclass correlation coefficients for the multilevel model for average STRAIN. \label{tab:icc}}
\centering
\begin{tabular}{cccccc}
\hline
\text{ } & Pass rusher & Pass blocker & Defensive team & Offensive team & Residual \\ 
\hline
ICC & 0.0365 & 0.0098 & 0.0134 & 0.0143 & 0.9260 \\ 
\hline
\end{tabular}
\end{table}

\begin{figure}

{\centering \includegraphics{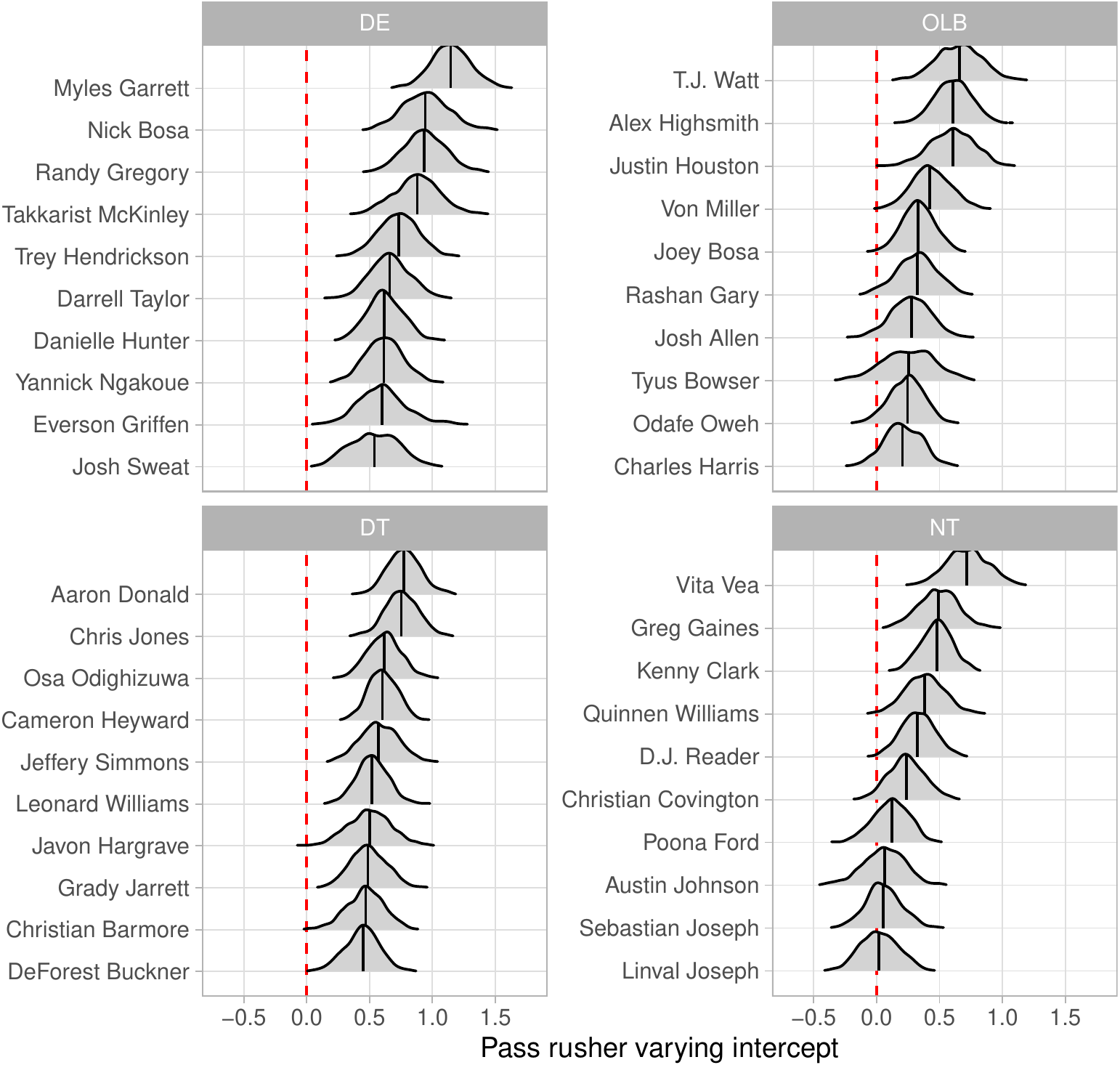} 

}

\caption{ Distributions of player effects (obtained from 1,000 bootstrap samples) for top 10 pass rushers in each position (defensive end, outside linebacker, defensive tackle, nose tackle) by median varying intercept.}\label{fig:fig_rankings_boot}
\end{figure}

Aside from adjusting for opposition, our multilevel model allows us to
provide some notion of uncertainty about the pass rusher's effect on the
play level STRAIN. Figure \ref{fig:fig_rankings_boot} displays the
varying intercept distributions for the top ten pass rushers for each
position. We obtain similar rankings based on the random intercepts in
comparison to the average STRAIN results in Tables \ref{tab:edge} and
\ref{tab:interior}, with Myles Garrett (defensive end), T.J. Watt
(outside linebacker), Aaron Donald (defensive tackle), and Vita Vea
(nose tackle) as the first-ranked pass rusher for their respective
position. Additionally, we observe different levels of variability
across these top defenders, with only a subset of players having
intercept distributions strictly above zero. This is not necessarily
surprising given the limited sample of data.

\hypertarget{sec:discussion}{%
\section{Discussion}\label{sec:discussion}}

In this work, we have proposed STRAIN---a simple and interpretable
statistic for evaluating pass rushers---with higher values corresponding
to greater pass rushing ability. STRAIN is a model-free metric which can
easily be integrated into any data pipeline without much computational
cost. Visualizations of STRAIN can be useful and intuitive for coaches
and broadcasters in various aspects ranging from gameplan preparation to
in-game real-time play analysis. We demonstrate that a pass rusher's
average STRAIN is both stable and more predictive of future pressure
events than using the player's previously observed pressure events. This
is analogous to the use of exit velocity in baseball, a predictive
measurement that avoids the noisy nature of observed outcomes,
emphasizing how the opportunity of working with player tracking data can
lead to the development of insightful metrics in American football.
Through multilevel modeling, we observe that the pass rusher explains
more variation in their play-level STRAIN, followed by the possession
team and pass rushing team, and finally the pass rusher's assigned
blocker. This is an intuitive result, consistent with previous
literature on pass rushing and defensive pressure.

Our multilevel model, however, is subject to several limitations. We
only account for players directly involved as the pass rusher or nearest
blocker, on top of team-level effects. The nested structure of our model
enforces positive dependence between pass rushers on the same team,
which may not be true. For instance, if one defender is known to be an
elite pass rusher then the opposing team may focus their blocking
efforts on this player. This could leave the path open for another
player to rush the quarterback with ease, resulting in higher STRAIN
values. As shown in Figure \ref{fig:fig_rankings_boot}, the first- and
fourth-ranked defensive ends Myles Garrett and Takkarist McKinley were
teammates on the Cleveland Browns during the first eight weeks of the
2021 NFL season. McKinley is a surprising name on our list and we
suspect that his high rank is mostly due to playing on the same
defensive unit with Garrett, who is highly-regarded as a great pass
rusher. In order to capture this type of behavior, we could consider
modeling an aggregate STRAIN across pass rushers with a regularized
adjusted plus-minus (RAPM) regression approach. The use of RAPM
techniques has been successfully demonstrated in American football by
\citet{sabin2021estimating}, which accounts for all players on the field
with Bayesian hierarchical models. We leave this type of analysis for
future work, which will require careful consideration of available
informative priors \citep{matano2023augmenting}.

In addition, we believe there are other concepts in materials science
that could be applied to evaluating pass rushing and pass blocking in
American football. One potential idea is to consider a quantity called
stress, which measures force over an area. By definition, force is the
product of mass and acceleration, meaning this quantity would take into
account the physical size of a pass rusher in the computation of force.
We could then divide this force over the ``area'' of the pocket formed
by pass blockers to compute stress.

Moreover, although we focus solely on pass rushers in this paper, STRAIN
can also be applied to the assessment of pass blockers as a unit. This
can be accomplished through the aforementioned RAPM regression or by
simply looking at quantities such as the total STRAIN or the maximum
STRAIN per frame aggregated across the entire play. It is also possible
to apply STRAIN to assess individual offensive pass blockers, provided
that a method of matching blockers to rushers is developed.
Furthermore, compared to existing metrics, STRAIN measures pass rush
effectiveness for every play continuously over time, which is at a much
more granular level than considering whether the play resulted in a
binary outcome such as a sack. Indeed, visualizations of STRAIN curves
across moments of time within plays reveal variability that simple
averaging may obscure. There is ample opportunity for working with the
complete STRAIN trajectories via temporal modeling and functional data
analysis techniques to better understand the impact of STRAIN on
offensive production in American football.

\hypertarget{acknowledgements}{%
\section*{Acknowledgements}\label{acknowledgements}}
\addcontentsline{toc}{section}{Acknowledgements}

We thank the organizers of the NFL Big Data Bowl 2023 for hosting the
competition and providing access to the data.

\newpage
\begin{center}
{\large\bf SUPPLEMENTARY MATERIAL}
\end{center}

All code related to this paper is available at
\url{https://github.com/getstrained/intro-strain}. The data provided by
the NFL Big Data Bowl 2023 is available at
\url{https://www.kaggle.com/competitions/nfl-big-data-bowl-2023/data}.

\bibliographystyle{chicago}
\renewcommand\refname{References}
\bibliography{bibliography.bib}

\begin{thebibliography}{}

\bibitem[\protect\citeauthoryear{Alamar and Goldner}{Alamar and
  Goldner}{2011}]{AlamarGoldner2011}
Alamar, B. and K.~Goldner (2011).
\newblock The blindside project: Measuring the impact of individual offensive
  linemen.
\newblock {\em Chance\/}~{\em 24}, 25--29.

\bibitem[\protect\citeauthoryear{Alamar and Weinstein-Gould}{Alamar and
  Weinstein-Gould}{2008}]{AlamarGould2008}
Alamar, B.~C. and J.~Weinstein-Gould (2008).
\newblock Isolating the effect of individual linemen on the passing game in the
  national football league.
\newblock {\em Journal of Quantitative Analysis in Sports\/}~{\em 4\/}(2).

\bibitem[\protect\citeauthoryear{Bates, M{\"a}chler, Bolker, and Walker}{Bates
  et~al.}{2015}]{bateslme4}
Bates, D., M.~M{\"a}chler, B.~Bolker, and S.~Walker (2015).
\newblock Fitting linear mixed-effects models using {lme4}.
\newblock {\em Journal of Statistical Software\/}~{\em 67\/}(1), 1--48.

\bibitem[\protect\citeauthoryear{Baumer, Matthews, and Nguyen}{Baumer
  et~al.}{2023}]{Baumer2023Big}
Baumer, B.~S., G.~J. Matthews, and Q.~Nguyen (2023).
\newblock Big ideas in sports analytics and statistical tools for their
  investigation.
\newblock {\em {WIREs} Computational Statistics\/}, e1612.

\bibitem[\protect\citeauthoryear{Burke}{Burke}{2018}]{Burke2018Created}
Burke, B. (2018).
\newblock {We created better pass-rusher and pass-blocker stats: How they
  work}.
\newblock ESPN.com.
\newblock [Accessed 16-May-2023].

\bibitem[\protect\citeauthoryear{Burke}{Burke}{2019}]{burke2019deepqb}
Burke, B. (2019).
\newblock Deepqb: deep learning with player tracking to quantify quarterback
  decision-making \& performance.
\newblock In {\em Proceedings of the 2019 MIT Sloan Sports Analytics
  Conference}.

\bibitem[\protect\citeauthoryear{Callister and Rethwisch}{Callister and
  Rethwisch}{2018}]{callister2018materials}
Callister, W.~D. and D.~G. Rethwisch (2018).
\newblock {\em Materials science and engineering: an introduction\/} (10 ed.).
\newblock Hoboken, NJ: John Wiley \& Sons, Inc.

\bibitem[\protect\citeauthoryear{Chu, Reyers, Thomson, and Wu}{Chu
  et~al.}{2020}]{chu2020route}
Chu, D., M.~Reyers, J.~Thomson, and L.~Y. Wu (2020).
\newblock Route identification in the national football league.
\newblock {\em Journal of Quantitative Analysis in Sports\/}~{\em 16\/}(2),
  121--132.

\bibitem[\protect\citeauthoryear{Deshpande and Evans}{Deshpande and
  Evans}{2020}]{deshpande2020expected}
Deshpande, S.~K. and K.~Evans (2020).
\newblock Expected hypothetical completion probability.
\newblock {\em Journal of Quantitative Analysis in Sports\/}~{\em 16\/}(2),
  85--94.

\bibitem[\protect\citeauthoryear{Dutta, Yurko, and Ventura}{Dutta
  et~al.}{2020}]{dutta2020unsupervised}
Dutta, R., R.~Yurko, and S.~L. Ventura (2020).
\newblock Unsupervised methods for identifying pass coverage among defensive
  backs with nfl player tracking data.
\newblock {\em Journal of Quantitative Analysis in Sports\/}~{\em 16\/}(2),
  143--161.

\bibitem[\protect\citeauthoryear{Eager and Chahrouri}{Eager and
  Chahrouri}{2018}]{Eager2018nfl}
Eager, E. and G.~Chahrouri (2018).
\newblock {Edge vs Interior: Which pass-rusher reigns supreme}.
\newblock PFF.com.
\newblock [Accessed 16-May-2023].

\bibitem[\protect\citeauthoryear{Hermsmeyer}{Hermsmeyer}{2021}]{Hermsmeyer2021nfl}
Hermsmeyer, J. (2021).
\newblock {The NFL Has A New Way To Measure The Explosiveness Of Pass Rushers}.
\newblock FiveThirtyEight.com.
\newblock [Accessed 16-May-2023].

\bibitem[\protect\citeauthoryear{Howard, Blake, Patton, Lopez, Bliss, and
  Cukierski}{Howard et~al.}{2022}]{Howard2023NFL}
Howard, A., A.~Blake, A.~Patton, M.~Lopez, T.~Bliss, and W.~Cukierski (2022).
\newblock {NFL Big Data Bowl 2023}.

\bibitem[\protect\citeauthoryear{Kovalchik}{Kovalchik}{2023}]{Kovalchik2023Player}
Kovalchik, S.~A. (2023).
\newblock Player tracking data in sports.
\newblock {\em Annual Review of Statistics and Its Application\/}~{\em
  10\/}(1), 677--697.

\bibitem[\protect\citeauthoryear{Linsey}{Linsey}{2022}]{Linsey2022nfl}
Linsey, B. (2022).
\newblock {2022 NFL interior defender rankings and tiers}.
\newblock PFF.com.
\newblock [Accessed 16-May-2023].

\bibitem[\protect\citeauthoryear{Lopez}{Lopez}{2020}]{lopez2020bigger}
Lopez, M.~J. (2020).
\newblock Bigger data, better questions, and a return to fourth down behavior:
  an introduction to a special issue on tracking datain the national football
  league.
\newblock {\em Journal of Quantitative Analysis in Sports\/}~{\em 16\/}(2),
  73--79.

\bibitem[\protect\citeauthoryear{Macdonald}{Macdonald}{2020}]{Macdonald2020Recreating}
Macdonald, B. (2020).
\newblock Recreating the {Game}: Using {Player} {Tracking} {Data} to {Analyze}
  {Dynamics} in {Basketball} and {Football}.
\newblock {\em Harvard Data Science Review\/}~{\em 2\/}(4).

\bibitem[\protect\citeauthoryear{Matano, Richardson, Pospisil, Politsch, and
  Qin}{Matano et~al.}{2023}]{matano2023augmenting}
Matano, F., L.~Richardson, T.~Pospisil, C.~A. Politsch, and J.~Qin (2023).
\newblock Augmenting adjusted plus-minus in soccer with fifa ratings.
\newblock {\em Journal of Quantitative Analysis in Sports\/}~{\em 19\/}(1),
  43--49.

\bibitem[\protect\citeauthoryear{Monson}{Monson}{2022}]{Monson2022nfl}
Monson, S. (2022).
\newblock {2022 NFL Edge Rusher Rankings and Tiers}.
\newblock PFF.com.
\newblock [Accessed 16-May-2023].

\bibitem[\protect\citeauthoryear{{NFL Football Operations}}{{NFL Football
  Operations}}{2023a}]{nfl2023big}
{NFL Football Operations} (2023a).
\newblock {Big Data Bowl}.
\newblock NFL.com.
\newblock [Accessed 16-May-2023].

\bibitem[\protect\citeauthoryear{{NFL Football Operations}}{{NFL Football
  Operations}}{2023b}]{nfl2023ngs}
{NFL Football Operations} (2023b).
\newblock {NFL Next Gen Stats}.
\newblock NFL.com.
\newblock [Accessed 16-May-2023].

\bibitem[\protect\citeauthoryear{{R Core Team}}{{R Core
  Team}}{2023}]{R2023Language}
{R Core Team} (2023).
\newblock {\em R: A Language and Environment for Statistical Computing}.
\newblock Vienna, Austria: R Foundation for Statistical Computing.

\bibitem[\protect\citeauthoryear{Reyers and Swartz}{Reyers and
  Swartz}{2021}]{reyers2021quarterback}
Reyers, M. and T.~B. Swartz (2021).
\newblock Quarterback evaluation in the national football league using tracking
  data.
\newblock {\em AStA Advances in Statistical Analysis\/}, 1--16.

\bibitem[\protect\citeauthoryear{Sabin}{Sabin}{2021}]{sabin2021estimating}
Sabin, R.~P. (2021).
\newblock Estimating player value in american football using plus--minus
  models.
\newblock {\em Journal of Quantitative Analysis in Sports\/}~{\em 17\/}(4),
  313--364.

\bibitem[\protect\citeauthoryear{Sterken}{Sterken}{2019}]{sterken2019routenet}
Sterken, N. (2019).
\newblock Routenet: a convolutional neuralnetwork for classifying routes.
\newblock {\em NFL Big Data Bowl\/}.

\bibitem[\protect\citeauthoryear{Wolfson, Addona, and Schmicker}{Wolfson
  et~al.}{2017}]{Wolfson2017Forecasting}
Wolfson, J., V.~Addona, and R.~Schmicker (2017).
\newblock Forecasting the performance of college prospects selected in the
  national football league draft.
\newblock In {\em Handbook of Statistical Methods and Analyses in Sports}, pp.\
   137--163. Boca Raton, FL: CRC Press.

\bibitem[\protect\citeauthoryear{Yurko, Matano, Richardson, Granered, Pospisil,
  Pelechrinis, and Ventura}{Yurko et~al.}{2020}]{yurko2020going}
Yurko, R., F.~Matano, L.~F. Richardson, N.~Granered, T.~Pospisil,
  K.~Pelechrinis, and S.~L. Ventura (2020).
\newblock Going deep: models for continuous-time within-play valuation of game
  outcomes in american football with tracking data.
\newblock {\em Journal of Quantitative Analysis in Sports\/}~{\em 16\/}(2),
  163--182.

\bibitem[\protect\citeauthoryear{Yurko, Ventura, and Horowitz}{Yurko
  et~al.}{2019}]{yurko2019nflwar}
Yurko, R., S.~Ventura, and M.~Horowitz (2019).
\newblock {nflWAR}: a reproducible method for offensive player evaluation in
  football.
\newblock {\em Journal of Quantitative Analysis in Sports\/}~{\em 15\/}(3),
  163--183.

\end{thebibliography}

\end{document}